\documentclass[11pt]{article}

\usepackage{amsmath,amssymb,mathtools,amsfonts,epsfig,graphicx,euscript}
\usepackage{amsmath,amssymb}
\newcommand{\nn}{\nonumber\\}

\newcommand{\bse}{\begin{subequations}}
\newcommand{\ese}{\end{subequations}}
\newcommand{\be}{\begin{equation}}
\newcommand{\ee}{\end{equation}}
\newcommand{\bea}{\begin{eqnarray}}
\newcommand{\eea}{\end{eqnarray}}
\newcommand{\ba}{\begin{array}}
\newcommand{\ea}{\end{array}}

\newcommand{\ie}{{\it i.e.}}

\begin{document}
\begin{titlepage}
\thispagestyle{empty}

\vspace{2cm}
\begin{center}
\font\titlerm=cmr10 scaled\magstep4 \font\titlei=cmmi10
scaled\magstep4 \font\titleis=cmmi7 scaled\magstep4 {
\Large{\textbf{Conductivity bound from dirty black holes}
\\}}

\vspace{1.5cm} \noindent{{
Kazem Bitaghsir Fadafan$^{}$\footnote{e-mail:bitaghsir@shahroodut.ac.ir },
}}\\
\vspace{0.8cm}

{\it ${}$Physics Department, Shahrood University of
Technology,\\ P.O.Box 3619995161 Shahrood, IRAN\\}

\vspace*{.4cm}

\end{center}
\vskip 2em

\begin{abstract}
We propose a lower bound of the dc electrical conductivity in strongly disordered, strongly interacting quantum field theories using holography. We study linear response of black holes with broken translational symmetry in Einstein-Maxwell-dilaton theories of gravity. Using the generalized Stokes equations at the horizon, we derive the lower bound of the electrical conductivity for the dual two dimensional disordered field theory.

\end{abstract}

\end{titlepage}

\section{Introduction}
Based on experimental results, study of some quantum systems in condensed matter theory such as strange metals and cold atomic phase need to consider strongly interacting many-body quantum physics \cite{Adams:2012th}. Quantum critical physics has important role in describing macroscopic observables in these systems \cite{Sachdev:2011cs}. One of the important observables is the electrical conductivity at finite density and disorder. To study such strongly interacting systems, the usual theoretical methods are not efficient and one should use new methods which are based on the nonperturbative approaches.

Using the gauge-string correspondence is a new tool for studying the transport coefficients in a strongly interacting systems. In this way, the quantum dynamics is encoded in the classical gravity in an asymptotically AdS spacetime. To consider finite temperature systems, one should add a black hole in the bulk space. Such systems in the condensed matter physics have large N matrix degrees of freedom. In this paper we are interested in studying the electrical DC conductivity at finite density and disorder. It is well known that a charged fluid with Galilean-invariant symmetry has infinite electrical conductivity. It is consequence of momentum conservation in the theory. Hence, all currents will have a non-zero conserved momentum and so they will not decay. However, recent studies from numerical holography in systems with breaking translation symmetry exhibits finite conductivity \cite{Horowitz:2012ky, Horowitz:2012gs,Horowitz:2013jaa,Chesler:2013qla}. Recent studies of the DC conductivity have been done for a charged fluid on the black hole horizon \cite{Donos:2015gia,Donos:2012js,Donos:2014gya,Banks:2015wha}. One should notice that it is related to the near horizon geometry of black holes.

To study strongly interacting quantum disorder systems from holography, massive gravity has been studied in \cite{Baggioli:2014roa,Alberte:2015isw}.
In such systems momentum relaxation is achieved even though translations are not explicitly broken in the bulk of the geometry. One important prediction of these models is that the disorder alone does not derive a metal-insulator transition. In mean-field disorder systems, insulators are not disorder-driven. The formation of a gap which should be proportional to the amount of disorder or additional localized features was studied in \cite{Arean:2015sqa}. In such studies the effects of disorder on a holographic superconductor has been investigated by introducing a random chemical potential on the boundary \cite{Arean:2014oaa,Arean:2013mta}. It was shown that increasing disorder leads to increasing the superconducting order and subsequently to the transition to a metal.

Recently, absence of disorder-driven metal-insulator transitions has been studied from holography \cite{Grozdanov:2015qia}. It was found that the electric DC conductivity of simple holographic disorder systems is bounded by the following universal value
\be \sigma \geq 1.\ee
This quantity $\sigma$ is measured in units of $\frac{e^2}{\hbar}$, with $e$ the $U(1)$ charge of the carriers. \footnote{We consider $\frac{e^2}{\hbar}=1$.} This bound means that one can not get an insulating phase by disorder-driven. They consider Einstein-Maxwell theory on $AdS_4$ without any free parameter and show that the bound does not depend on the temperature or fluctuations in the charge density. The simple holographic model in \cite{Grozdanov:2015qia} means that there is no coupling between scalar dilaton $\phi$ to the Maxwell field also no additional charged fields in the bulk of the theory.

In this paper we extend the results of \cite{Grozdanov:2015qia} and obtain the bound for DC conductivity of disordered Einstein-Maxwell-dilaton (EMD) holographic systems. We bound $\sigma$ in a relativistic theory which is dual to a disorder black hole in EMD geometry and in two spatial dimension. In such holographic  models the field theory is deformed by a charge-neutral relevant scalar operator. In general EMD holographic models, the DC conductivity $\sigma$ can be either metallic or insulating \cite{Rangamani:2015hka,Donos:2014uba,Gouteraux:2014hca}. Such models are interesting to study the strange metal phase of high temperature superconductors.

We have studied before two important properties of the strange metals, the Ohmic resistivity and the inverse Hall angle, in the presence of finite-coupling corrections in \cite{Fadafan:2012hr}. In this study, we considered the AdS spacetime in the light-cone frame. This frame could be used to study physical properties of strange metals \cite{Kim:2010zq}. The electrical DC conductivity of massive $\mathcal{N} = 2$ hypermultiplet fields at finite temperature and in an $\mathcal{N} = 4$ $SU(N_c)$ super-Yang-Mills theory in the large $N_c$ and finite-coupling correction was studied in \cite{AliAkbari:2010av}.\\

This paper is organized as follows. In the next section we review the EMD gravity in \cite{Banks:2015wha} and study the charged horizon fluid. We also define heat and electric currents in the horizon. In section two we use variational method and derive the bound on the conductivity. In the last section we discuss and summarize our results. %

\section{Dirty black holes}
The EMD holographic theories in $D$ spacetime dimensions have been studied recently in \cite{Banks:2015wha}.  In the following the case of $AdS_4$ has been considered, also some notations of \cite{Banks:2015wha} has been changed. In the following we discuss the general class of holographic lattices. Periodic lattices with chemical potential and real scalar field have been studied in \cite{Horowitz:2012ky,Ling:2013nxa}. Also Q-lattice models with more scalar fields were studied in \cite{Donos:2013eha,Donos:2014cya,Kiritsis:2015oxa}. The helical lattices were studied in \cite{Donos:2012js,Donos:2014gya,Donos:2014oha}. Study of holographic lattice models in the presence of magnetic fields has been done in \cite{Blake:2015ina, Amoretti:2015gna,Donos:2015bxe}. The generalization of the results of this section in the presence of magnetic fields has been done in \cite{Donos:2015bxe}.

The action is given by
\begin{align}\label{action}
S=\int d^4 x \sqrt{-g}\,\left(R-V(\phi)-\frac{Z(\phi)}{4}\,F^{2}-\frac{1}{2}\left(\partial\phi \right)^2\right)\,.
\end{align}
Here, $F$ is the Maxwell field strength of $A$ which is dual of a global $U(1)$ gauge field. There is also an operator dual to the scalar dilaton field $\phi$. One should assume $V(\phi=0)=-6$,$V'(\phi=0)=0$ and $Z(\phi=0)=1$ to find a unit radius $AdS_4$ geometry. We have set the $AdS_4$ radius to unity, as well as setting $16 \pi G=1$.  The equations of motion are given by
\begin{align}\label{metric}
R_{\mu\nu}-\frac{V(\phi)}{2}g_{\mu\nu}-\frac{1}{2}\partial_\mu\phi\partial_\nu\phi-\frac{1}{2}Z(\phi)\left(
F_{\mu\rho}F_{\nu}{}^{\rho}- \frac{1}{4}g_{\mu\nu}\,F^2\right)&=0\,,\nn
 \nabla_{\mu}\left[Z(\phi)F^{\mu\nu}\right]&=0\,,\nn
 \nabla^2\phi-V'(\phi)-\frac{1}{4}Z'(\phi)F^2&=0\,.
 \end{align}
We consider a general static electrically charged black hole geometry as
 \begin{align}\label{metric}
ds^{2}=-U(r)V(r,\textbf{x})\,dt^{2}+\frac{W(r,\textbf{x})}{U(r)}\,dr^{2}+G_{ij}dx_idx_j\,,
\end{align}
with a gauge-field $A$ given by
\be A=a_t(r,\textbf{x})\,dt=\Phi(r,\textbf{x})\,dt\,,\ee
Here $G_{ij}$ is a metric on a two dimensional manifold at fixed $r$. The holographic direction is denoted by $r$ and the boundary field theory directions are $(t,\vec{x})$. A connected black horizon is located at $r=0$. The dual field theory temperature $T$ is given by the Hawking temperature of black hole. In the UV boundary condition, as $r\to\infty$, the solutions are taken to approach $AdS_4$ spacetime.  The gauge field also goes to the spatially dependent chemical potential $\mu(x)$ in the boundary. To consider periodic lattices, one assumes periodic conditions for fields in the boundary. The period in spatial directions is denoted by $L_i$ and the geometry at fixed $r$ parameterizes a torus with period $x_i\sim x_i+L_i$ also the black hole horizon has the same topology. With using Kruskal coordinate $v=t+\frac{ln\,r}{4\pi T}+\ldots$, one finds the near horizon expansions of the metric functions and fields as
\begin{align}\label{nhorizon}
U\left(r\right)&=4\pi\,T\,r+U^{(1)}\,r^2+\dots\,,\nn
a_{t}(r,x)&=a^{(0)}_{t}W^{(0)}\left(x\right)r+a^{(1)}_{t}\left(x\right)r^2+\dots\,,\nn
W(r,x)&=W^{(0)}\left(x\right)+W^{(1)}\left(x\right)r+\dots\,,\nn
V(r,x)&=W^{(0)}\left(x\right)+V^{(1)}\left(x\right)r+\dots\,,\nn
G_{ij}&=G_{ij}^{(0)}+G_{ij}^{(1)}r+\dots\,,\nn
\phi(x)&=\phi^{(0)}(x)+r\phi^{(1)}(x)+\dots\,,
\end{align}
One should notice that $W^{(0)}(x)=V^{(0)}(x)$.

The electric charge density at the horizon is $\rho_h=\sqrt{-g}Z(\phi)F^{tr}|_h=\sqrt{-g_0}a^{(0)}_{t}Z^{(0)}$ where $Z^{(0)}\equiv Z(\phi^{(0)}(x))$ and $a^{(0)}\equiv a(\phi^{(0)}(x))$. The scalar dilaton value at the horizon is denoted as $\phi^{(0)}$. Henceforth, we use such notation for values of parameters near the horizon.

By turning on electric current $E$ and temperature gradient $\zeta$ on the geometry at fixed $r$, the black hole will response \cite{Donos:2014cya}. One should assume $E_i=E_i(x)$, $\zeta_i=\zeta_i(x)$ and use the appropriate linear perturbations $\delta g_{\mu\nu}$,$\delta a_\mu$, $\delta\phi$ which are functions of $(r,x^i)$ \cite{Banks:2015wha}. At the black hole horizon, the leading order terms are
\begin{align}\label{linear1}
\delta g_{tt}&\rightarrow U\left(r \right)\,\delta g^{(0)}_{tt}\left(x\right),\quad
\delta g_{rr}\rightarrow \frac{\delta g_{rr}^{(0)}\left(x\right)}{U},\quad
\delta g_{ij}\rightarrow \delta g_{ij}^{(0)}\left(x\right),
\quad
\delta g_{tr}\rightarrow \delta g_{tr}^{(0)}\left(x\right)\,,\nn
\delta g_{ti}&\rightarrow \delta g_{ti}^{(0)}\left(x\right)-\,V(r,\textbf{x}) U(r)\zeta_{i}\,\frac{\ln r}{4\pi T},\quad
\delta g_{ri}\rightarrow \frac{1}{U}\,\left( \delta g_{ri}^{(0)}\left(x\right) \right)\,,\nn
\delta a_{t}&\rightarrow \delta a_{t}^{(0)}\left(x\right),\quad
\delta a_{r}\rightarrow \frac{1}{U}\,\left(\delta a_{r}^{(0)}\left(x\right)\right)\,,\quad
\delta a_{i}\rightarrow \frac{\ln{r}}{4\pi T}(-E_i+a_t(r,\textbf{x})\zeta_i)\,,
\end{align}
with the following constraints on the leading order terms of $x$ as
\begin{align}
&\delta g_{tt}^{(0)}+\delta g_{rr}^{(0)}-2\,\delta g_{rt}^{(0)}=0,\qquad \delta g^{(0)}_{ri}=\delta g^{(0)}_{ti},\qquad
\delta a_{r}^{(0)}=\delta a_{t}^{(0)}\,.
\end{align}

\subsection{The electric and heat current }
The bulk electric current density is defined as
\begin{align}
J^i=\sqrt{-g}Z(\phi)F^{ir}\,.
\end{align}
At linearized order for the perturbed black holes, one finds
\begin{align}\label{jexpression}
J^i=\frac{\sqrt{g}G^{ij}}{(VW)^{1/2}}VUZ(\phi)\Bigg(\partial_ja_t \frac{\delta g_{rt}}{VU}-\partial_r a_t(\frac{\delta g_{jt}}{VU}-\frac{tVU\zeta_j}{VU})
+\partial_j\delta a_r-(\partial_r\delta a_j+t \partial_r a_t\zeta_j)\Bigg)\,,
\end{align}
and it is clearly seen that the time-dependence drops out.

The heat current density is given by $Q$ and at linearized order one finds it in terms of the perturbed black holes metrics as
\begin{align}
Q^i=\frac{V^{3/2}U^2}{W^{1/2}}\sqrt{g}G^{ij}\left(\partial_r\left(\frac{\delta g_{jt}}{VU}\right)-\partial_j\left(\frac{\delta g_{rt}}{VU}\right)
\right)-a_t J^i\,.
\end{align}
At linearized order, one finds that $\partial_iQ^i=0$.

The expressions for the electric and the heat current densities at the black hole horizon is given by:
\begin{align}
J^i_{(0)}&\equiv\left.J^i\right|_h =Z^{(0)}\sqrt{g_{(0)}}G^{ij}_{(0)}\left(\left(\partial_j\delta a_t^{(0)}+E_j\right)-a^{(0)}_t\delta g^{(0)}_{jt}\right)\,,\nn
Q^i_{(0)}&\equiv\left.Q^i\right|_h =-4\pi T\sqrt{g_{(0)}}G^{ij}_{(0)}\delta g^{(0)}_{jt}\,.
\end{align}
One obtains that they are conserved currents as $\partial_iJ^i_{(0)}=0$ and $\partial_iQ^i_{(0)}=0$.
\section{Stokes equations at the horizon and bound of electrical conductivity}
In this section we use the variational techniques proposed in \cite{Lucas:2015lna} and study the lower bound on the DC conductivity. This method has been used in \cite{Ikeda:2016rqh} to find upper and lower bounds on the electrical conductivity of holographic disorder probe brane models at finite temperature. This approach is analogous to study of the resistance of a disordered resistor network when one runs any set of trial currents. One finds an upper bound of the resistivity by calculating the dissipated power.

Using the linearized perturbations, one finds a closed system of equations at the black hole horizon. They will give linear system of partial differential equations by introducing new notations as
\begin{align}
v_{i}\equiv -\delta g_{it}^{(0)},\qquad w\equiv \delta {a}_{t}^{(0)},\qquad p\equiv -4 \pi T\frac{\delta g_{rt}^{(0)}}{V^{(0)}}-\delta g_{it}^{(0)}g^{ij}_{(0)}\nabla_{j}\,\ln V^{(0)}\,.
\end{align}

The linear system of partial differential equations at the horizon fluid are
\begin{align}
\nabla_{i} v^{i}&=0\,, \label{stokes1}\\
\mathcal{J}^i=\left(Z^{(0)}\nabla^i w+v^{i}\,{Z^{(0)}a_{t}^{(0)}}+Z^{(0)} E^{i}\right),\,\,\,\,\,\,\,\,\,\nabla_i\mathcal{J}^i&=0,\label{stokes2}\\
-2\,\nabla^{i}\nabla_{\left( i \right. }v_{\left. j\right)}-{Z^{(0)}a_{t}^{(0)}}\nabla_j w
+\nabla_j\phi^{(0)}\nabla_i\phi^{(0)}v^{i}
+\nabla_{j}\,p-4\pi T\,\zeta_{j}
-{Z^{(0)}a_{t}^{(0)}} E_j&=0\,,\label{stokes}
\end{align}
Here, all indices are being lowered and raised with respect to the black hole metric, $G_{ij}^{(0)}$, also the covariant derivative taken with it. These equations on the curved black hole horizon are a generalization of the forced Stokes equations for a charged fluid. It is interesting that considering the scalar field $\phi$ leads to new term in the Stokes equations as $\nabla_j\phi^{(0)}\nabla_i\phi^{(0)}v^{i}$. It can be interpreted as a friction term.

The first two equations give us the conserved currents at the horizon fluid.  They are denoted as $v^i$ and $\mathcal{J}^i$. We normalize them so that  $Q^i_{(0)}, J^i_{(0)}$ are heat current and electric current densities at the horizon fluid, respectively:
\begin{align}\label{g1}
J^i_{(0)}&=\sqrt{g_{(0)}}G^{ij}_{(0)}\mathcal{J}_j\,,\nn
Q^i_{(0)} &=4\pi T\,\sqrt{g_{(0)}}v^i\,.
\end{align}

\subsection{Bound of electrical conductivity}

Using Stokes equations, one finds that heat current  $Q^i_{(0)}$ and electric current $J^i_{(0)}$ are conserved currents. Also, they depends on the coordinates of the horizon. First we normalize these quantities by spatially averaging them over the horizon quantities as
\begin{align}
J^i_{charge}&=\mathbb{E}[J^i_{(0)}]\,,\nn
Q^i_{heat} &=\mathbb{E}[Q^i_{(0)}]\,.
\end{align}
Here we used $E[\mathcal{O}]=\frac{1}{L^2}\int d^2x\,\mathcal{O}$.

The dissipated power can be written as
\be P= E _i J^i_{charge}+ \xi_i Q^i_{heat}\ee
By employing Stokes equations, one finds
\bea
P= \mathbb{E}\left[2\sqrt{g}\,\nabla^{(i}v^{j)}\nabla_{(i}v_{j)} + \sqrt{g} \left(v^i \nabla_i \phi\right)^2
+\sqrt{g} Z^{(0)}\left(a_t v_i -\frac{\mathcal{J}_i}{Z^{(0)}} \right) \left(a_t v^i -\frac{\mathcal{J}^i}{Z^{(0)}} \right)\right],\nonumber  \label{power}
\eea
Following proposal of \cite{Lucas:2015lna}, we consider $P$ as a functional conserved currents $v^i$ and $\mathcal{J}^i$, \ie $P[v^i,\mathcal{J}^i]$. Thus for arbitrary conserved and periodic set of charge and heat currents in the spatial direction $v^i$ and $\mathcal{J}^i$, we divide them as follows:
\begin{equation}
\mathcal{J}^i = \bar{\mathcal{J}}^i  + \hat{\mathcal{J}}^i,\,\,\,\,\,\,v^i = \bar{v}^i  + \hat{v}^i
\end{equation}
here $(\bar{\mathcal{J}},\bar{v})$ means the true solution of the system of hydrodynamic equations with appropriate boundary conditions as they are shown in eqs. (\ref{stokes1}), (\ref{stokes2}) and (\ref{stokes}). Also $(\hat{\mathcal{J}},\hat{v})$ implies the deviations. Now we expand out $P[v^i,\mathcal{J}^i  ]$ as
\be
P[v^i,\mathcal{J}^i  ]=P[\bar{v}^i  + \hat{v}^i,\bar{\mathcal{J}}^i  + \hat{\mathcal{J}}^i]=P[\bar{v}^i,\bar{\mathcal{J}}^i ]+P[\hat{v}^i, \hat{\mathcal{J}}^i]+2\mathcal{P}
\ee
Here
\be
\mathcal{P}= 2\,\mathbb{E}[\sqrt{g}\,\nabla^{(i}\bar{v}^{j)}\nabla_{(i}\hat{v}_{j)}]+ \mathbb{E}[\sqrt{g} \left(\bar{v}^i \nabla_i \phi\right)\left(\hat{v}^i \nabla_i \phi\right)]+\mathbb{E}[\sqrt{g} Z^{(0)}\left(a_t \bar{v}_i -\frac{\bar{\mathcal{J}}_i}{Z^{(0)}} \right) \left(a_t \hat{v}^i -\frac{\hat{\mathcal{J}}^i}{Z^{(0)}} \right)]
\ee
Using the current equation and integrating by parts, one finds that
\be
\mathcal{P}=0.
\ee
Also $ P[v^i,\mathcal{J}^i], P[\bar{v}^i,\bar{\mathcal{J}}^i], P[\hat{v}^i,\hat{\mathcal{J}}^i]\geq 0$. As a result, we find that
\be P[v^i,\mathcal{J}^i]\geq P[\bar{v}^i,\bar{\mathcal{J}}^i]\label{P1}\ee

In the absence of heat current the dissipated power is
\be
P=\frac{J_{charge}^2}{\sigma}.
\ee
here we have considered $J_{charge}^i=\sigma E^i$.
We normalize the electric current at the horizon as
\be
J_{charge}^2=\mathbb{E}[\sqrt{g} \mathcal{J}^i\mathcal{J}_i].
\ee
Hence using the above statements, one finds the electrical conductivity bound.
We consider a simple guess for arbitrary currents which corresponds to strong momentum relaxation as
\be v^i=0,\,\,\,\,\,\,\, \mathcal{J}^i=\delta^i_x, \ee
It means that the metal behaves in the sector of diffusive.
The dissipated power for our guess would be
\be
P[0,\mathcal{J}^i]=\mathbb{E}[\frac{\sqrt{g}\mathcal{J}^i\mathcal{J}_i}{Z^{(0)}}]=\mathbb{E}[1/Z^{(0)}]J_{charge}^2,
\ee
Here, we have used normalization in eq.(26). Also when $\phi$ changes on the horizon, we should consider spatial averaging as $\mathbb{E}[1/Z^{(0)}]$. Using the inequality in eq. (\ref{P1}), we find
\be
\frac{J_{charge}^2}{\sigma} \leq \mathbb{E}[1/Z^{(0)}]J_{charge}^2,
\ee

Then one finds the new lower bound on $\sigma$ in the EMD theories as
\be
\sigma \geq \frac{1}{\mathbb{E}[1/Z^{(0)}]}.\label{bound}
\ee
 Without scalar field, $Z(\phi)=1$ and in the case of simple holographic models one obtains the lower bound in \cite{Grozdanov:2015qia}.
\section{Discussion }
In this paper we have studied lower bound of the electrical conductivity in strongly coupled disordered quantum field theories using holography. We studied linear response of black holes with broken translational symmetry in the presence of scalar dilaton field.

As it was proposed in \cite{Grozdanov:2015qia}, the lower bound on the electrical conductivity of simple holographic models does not hold in EMD theories, when $F^2$ couples to the dilaton. We discussed how to analytically study the bound of the conductivity in such holographic models in terms of black hole horizon data of EMD geometries. We used the generalized Stokes equations at the horizon and derived the lower bound of the dc electrical conductivity for the dual two dimensional disordered field theory.

We have used the variational techniques proposed in \cite{Lucas:2015lna}. Such study has been done in the case of thermal transport in strongly disordered and strongly interacting quantum field theories in \cite{Grozdanov:2015djs}. They derive bounds on thermal conductivity for the dual disordered field theories at finite temperature and in one and two spatial dimensions. It was shown that in one dimension it could be always non-zero.

It should be noticed that the insulating configurations in holographic EMD models have been built. In such models the metal-insulator transition is not driven by disorder, they arise because of the coupling of the scalar dilaton to the electromagnetic field. The dilaton will gap all the states in the system and the system would be an insulator. The lower bound of the DC electrical conductivity in such models has been found in \cite{Ge:2015fmu}. They consider EMD action with linear scalar fields and show that anisotropy is a fundamental factor in normal-state of high temperature superconductors. They also present a universal lower bound of dc electrical conductivity in anisotropic systems. The lower bound of eq. (\ref{bound}) matches to their result, too.

Regarding the main result in (\ref{bound}), it would be interesting to study metal-insulator transition by disorder in the presence of scalar field. In particular one has to prove that as a function of disorder strength $\mathbb{E}[1/Z^{(0)}]$ undergoes a qualitative change. However, it is quite challenging and requires solving the bulk PDEs with disordered boundary conditions. Then we interpret (\ref{bound}) only as a lower bound.

Recently, the lower bound in \cite{Grozdanov:2015qia} has been studied in \cite{Baggioli:2016oqk,Gouteraux:2016wxj} from effective holographic theories. For example, a minimal holographic model of a disorder-driven metal-insulator transition was introduced in \cite{Baggioli:2016oqk}. They do not introduce dilaton or additional sector in the model. They consider a Massive Gravity-Maxwell model with new effective interactions between the metric and electromagnetic field which are allowed when gravity is massive. It was shown that the conductivity can be very small in the insulating phase and does not obey any lower bound.

\section*{Acknowledgments}
I would like to thank Leopoldo A. Pando Zayas, Antonello Scardicchio and Daniel Arean for directing me on disorder-driven metal-insulator transitions during my stay at the International Center for Theoretical Physics (ICTP). Also many thank from Daniele Musso for interesting physics discussions. I thank the ICTP for warm hospitality. I am grateful to A. Lucas for useful discussions on the variational method.


\end{document}